\documentclass[preprint,showpacs,preprintnumbers]{revtex4}
\usepackage{amsmath}
\usepackage{graphicx}
\usepackage{dcolumn}
\usepackage{bm}
\usepackage{epsfig}
\usepackage[T1]{fontenc}
\usepackage{ae,aecompl}

\begin{document}

\title{Thermodynamic non-equilibrium effects in bubble coalescence: A
discrete Boltzmann study}
\author{Guanglan Sun$^{1,2}$, Yanbiao Gan$^{2}$, Aiguo Xu$^{3,4,5}$ \footnote{
Corresponding author. E-mail: Xu\_Aiguo@iapcm.ac.cn}, Yudong Zhang$^6$,
Qingfan Shi$^1$\footnote{Corresponding author. E-mail: qfshi123@bit.edu.cn}}
\affiliation{1, School of Physics, Beijing Institute of Technology, Beijing 100081, China
\\
2, Hebei Key Laboratory of Trans-Media Aerial Underwater Vehicle, School of
Liberal Arts and Sciences,\\
North China Institute of Aerospace Engineering, Langfang 065000, China \\
3, National Key Laboratory of Computational Physics, \\
Institute of Applied Physics and Computational Mathematics, P. O. Box
8009-26, Beijing 100088, P.R.China \\
4, State Key Laboratory of Explosion Science and Technology, Beijing
Institute of Technology, Beijing 100081, China \\
5, HEDPS, Center for Applied Physics and Technology, and College of
Engineering, Peking University, Beijing 100871, China \\
6, School of Mechanics and Safety Engineering, Zhengzhou University,
Zhengzhou 450001, China }
\date{\today }

\begin{abstract}
The Thermodynamic Non-Equilibrium (TNE) effects in the coalescing process of two initially static bubbles under thermal conditions are investigated by a Discrete Boltzmann Model (DBM). The spatial distributions of the typical none-quilibrium quantity, i.e., the Non-Organized Momentum Fluxes (NOMF) during evolutions are investigated in detail. The density-weighted statistical method is used to highlight the relationship between the TNE effects and the morphological or kinetics characteristics of bubble coalescence.
It is found that the $xx$-component and $yy$-component of NOMF are anti-symmetrical; the $xy$-component changes from an anti-symmetric internal and external double quadrupole structure to an outer octupole structure during the coalescing process.
More importantly, the evolution of the averaged $xx$-component of NOMF provides two characteristic instants, which divide the non-equilibrium process into three stages. The first instant corresponds to the moment when the mean coalescing speed gets the maximum and at this time the ratio of minor and major axes is about $1/2$. The second instant corresponds to the moment when the ratio of minor and major axes gets $1$ for the first time.
It is interesting to find that the three quantities, TNE intensity, acceleration of coalescence and negative slope of boundary length, show a high degree of correlation and attain their maxima simultaneously.
Surface tension and heat conduction accelerate the process of bubble coalescence while viscosity delays it. Both surface tension and viscosity enhance the global non-equilibrium intensity, whereas heat conduction restrains it.
These TNE features and findings present some new insights into the kinetics of bubble coalescence.
\end{abstract}

\maketitle
\preprint{APS/123-QED}

\section{Introduction}

Bubble coalescence is frequently encountered in many applications such as two-phase electrochemical systems \cite{RN1,RN2,RN3}, biological and pharmaceutical applications \cite{RN4,RN5,RN53Alexander,RN56Yeomans}, and boiling water-cooled packed bed reactors \cite{RN6,RN7,RN8}. In some cases, the coalescence of bubbles needs to be prevented, while in other cases, it must be promoted. Therefore, it is necessary to fundamentally understand the essence of bubble coalescence, including its basic dynamics phenomena, morphological characteristic, particularly the commonly neglected non-equilibrium effects and behaviors of the system, etc.

There have been many studies on the phenomenon and mechanism of bubble
coalescence through experiments \cite{RN9,RN10,RN11,RN12,RN48,RN14,RN15,RN16,doi:10.1063/1.4979087}, theoretical analyses \cite{RN10,RN17,RN18} and numerical simulations \cite{RN12,RN19,RN20,RN21,RN22,RN23,RN24,RN25,RN54Alexander,chai2011gas,Luoli2016lattice}. Stover, \emph{et al.} \cite{RN12} first investigated the turbulence characteristics of coalescence of bubbles with different sizes, viscosities, and surface tensions. They found that surface waves, started at the onset of coalescence, were superimposed on the motion of the bubbles. These waves are likely to enhance the mass transfer efficiency but have little effect on the overall dynamics. Then they simulated the dynamics of coalescence by solving the nonlinear Navier-Stokes equations, and their results were only partly consistent with the experiments due
to the failure of reflecting the actual initial state.
The approximate
analytic solution of the neck radius ${r_{n}} \sim B{t^{1/2}}$ was given in Refs. \cite{RN10,RN17,RN18,RN25}, the pre-factor $B$ was positively correlated with surface tension coefficient and negatively correlated with viscosity coefficient.  Zheng, \emph{et al.} \cite{RN20} developed an free-energy Lattice Boltzmann Method (LBM) to mick the interface. In the LBM, the interface is naturally captured. It does not require interface reconstruction as required by most traditional interface tracking methods, such as the volume of fluid method. They simulated the bubble coalescence at a high-density ratio using LBM \cite{RN21}. A similar approach, solving
Navier-Stokes equations and Cahn-Hilliard interface evolution equation by LBM, was used by Chen, \emph{et al.} to study the coalescence of unequal-size or equal-size bubbles \cite{RN22,RN23,RN24}. They presented power-law relations between the global coalescence time and size inequality, and the effect of the Ohnesorge (\rm{Oh}) number on those power-law relations.
They demonstrated that unequal bubbles coalesce faster than equal bubbles
 and clarified the relations between characteristic coefficients and the Oh number \cite{RN24}.
They presented a critical Oh value which can be used to predicate whether the post-fusion behavior with damped oscillations \cite{RN23}.

Despite these significant progresses to date, the Thermodynamic Non-Equilibrium (TNE) effects during bubble coalescence are rarely concerned.
However, as we will show in this paper, the TNE behaviors are of great importance for this dynamic and complex process, because the TNE effects enormously influence the morphological characteristic, decide the coalescing speed, macroscopic quantity and stress distributions, phase transformation speed, etc. The careful study of these behaviors is beyond the physical capability of the traditional hydrodynamic model. In this work, we resort to the recently developed discrete Boltzmann method/model (DBM).

DBM \cite{RN46,RN47,RN45,RN26,RN27,RN28,RN29,RN13} is a mesoscopic kinetic model. In 2012, Xu, \emph{et al.} \cite{RN46} pointed out that, under the framework of LBM and under the condition that does not use the non-physical Boltzmann equation and kinetic moments, the non-conservative moments of (${f_i} - f_i^{eq}$) can be used to describe how and how much the system deviates from the thermodynamic equilibrium, and to check corresponding effects due to deviating from the thermodynamic equilibrium. This was the starting point for the current DBM study. In 2015, Xu, \emph{et al.} \cite{RN47} proposed to open phase space using the non-conservative moments of (${f_i} - f_i^{eq}$) and describe the extent of TNE using the distance between a state point to the origin in the phase space or its sub-space. In 2018, Xu, \emph{et al.} \cite{RN45} further developed the non-conservative moment phase space description methodology. They proposed to use the distance $D$ between two state points to roughly describe the difference between the two states deviating from their thermodynamic equilibria, and the reciprocal of distance, $1/D$, is defined as a similarity of deviating from thermodynamic equilibrium. The mean distance during a time interval of $D$, $\bar D$, is used to roughly describe the difference between the two corresponding kinetic processes, and the reciprocal of $\bar D$, $1/\bar D$, is defined as a process similarity.
In 2021, Xu, \emph{et al.} \cite{RN13} extended the phase space description methodology to any system characteristics. Use a set of (independent) characteristic quantities to open phase space, and use this space and its sub-spaces to describe the system properties. A point in the phase space corresponds to a set of characteristic behavior of the system. Distance concepts in the phase space or its sub-spaces are used to describe the difference and similarity of behavior. Up to now, DBM has been used in various multiphase flow systems, such as hydrodynamic instabilities \cite{RN27,RN31,RN32,RN33,RN34,RN35}, compressible flows under impact \cite{RN26,RN27,RN28,RN29}, non-equilibrium combustion \cite{RN36,RN37,RN38}, non-equilibrium phase separation \cite{RN39,RN40,RN41,RN42}, and droplets collision \cite{RN43}. For example, Gan, \emph{et al.} \cite{RN39} used a DBM to study the Hydrodynamic Non-Equilibrium (HNE) and TNE effects in phase separation process. They defined TNE strength and discovered that the time evolution of the TNE intensity provides a convenient and efficient physical criterion to discriminate the stages of spinodal decomposition and domain growth. Lai, \emph{et al.} \cite{RN33} studied the effects of compressibility on Rayleigh-Taylor Instability (RTI) by DBM. It is found that the local TNE can be used to track the interfaces and discriminate between the two stages of the RTI effectively. Zhang, \emph{et al.} \cite{RN43} researched the droplet collisions by DBM on the basis of a discrete Enskog equation. They found that the mean strength of the Non-Organized Momentum Fluxes (NOMF) $\bar D_2^*$ was always prominently greater than that of the Non-Organized Energy Fluxes (NOEF) $\bar D_3^*$, and $\bar D_2^*$ can be used to identify the different stages of the collision process and to recognize different types of collisions. Zhang, \emph{et al.} \cite{RN40} analyzed entropy production associated with TNE of thermal phase separation. They obtained that NOMF and NOEF both directly contribute to entropy production.

In this paper, we systematically study the TNE characterizations of bubble coalescence via a compressible multiphase DBM. The rest of this paper is organized as follows. The physical model is presented in Sec. II; simulations and analysis of non-equilibrium characteristics during bubble coalescence and the effects of surface tension, viscosity, and heat conduction are presented in Sec. III; the conclusions are made in Sec. IV.

\section{Construction of DBM}

According to molecular kinetic theory, the evolution equation of molecular
velocity distribution function reads
\begin{equation}
{\frac{{\partial f}}{{\partial t}}}+\mathbf{v}\cdot {\frac{{\partial f}}{{%
\partial \mathbf{r}}}}+\mathbf{a}\cdot {\frac{{\partial f}}{{\partial
\mathbf{v}}}}={({\frac{{\partial f}}{{\partial t}}})_{c}},
\end{equation}%
where $f=f({\mathbf{r},\mathbf{v},t})$ is the molecular
velocity distribution function, $\mathbf{r}$, $\mathbf{v}$, and $\mathbf{a}$ are the position space coordinate, the velocity space coordinate, and the acceleration generated by the total extra force, respectively. $({\partial f}/{\partial t})_c$ is collision term, if the constraint
 $\int {\mathbf{\Psi }{({\partial f}/{\partial t})_c}}d\mathbf{v} = \int -\mathbf{\Psi }{({f-{f^{eq}}})}/{\tau} d\mathbf{v}$ is satisfied, the collision term can be linearized $({\partial f}/{\partial t})_c = -{({f-{f^{eq}}})}/{\tau}$, here $\mathbf{\Psi}=[1,\mathbf{v},\mathbf{vv}, \mathbf{vvv}]^{\rm T}$, $\tau$ is a relaxation time and ${f^{eq}}=\rho {({1/2\pi T})^{3/2}}\exp[-{{(\mathbf{v}-\mathbf{u})}^{2}}/2T] $ the Maxwellian distribution function in Bhatnagar-Gross-Krook (BGK) model.

To describe the nonideal gas effects, Gonnella, Lamura, and Sofonea (GLS)
improved the Watari-Tsutahara (WT) model \cite{RN44} by introducing an
appropriate force term in the right hand of Boltzmann-BGK equation
\begin{equation}
{\frac{{\partial f}}{{\partial t}}}+\mathbf{v}\cdot {\frac{{\partial f}}{{%
\partial \mathbf{r}}}}=-{\frac{1}{\tau }}(f-{f^{eq}})+I,
\end{equation}
with
\begin{equation}
I=-[A+\mathbf{B}\cdot (\mathbf{v}-\mathbf{u})+(C+{C_{q}}){(\mathbf{v}-%
\mathbf{u})^{2}}]{f^{eq}}.
\end{equation}%
Here
\begin{equation}
A=-2\left( {C+{C_{q}}}\right)T,
\end{equation}%
\begin{equation}
\mathbf{B}={\frac{1}{{\rho T}}}\nabla \cdot \left[ {\left( {{P^{\mathrm{CS}}-%
}\rho T}\right) \mathbf{I}+\mathbf{\Lambda }}\right],
\end{equation}%
\begin{eqnarray}
C &=&{\frac{1}{{2\rho {T^{2}}}}\left( {{P^{\mathrm{CS}}}-\rho T}\right)
\nabla \cdot \mathbf{u}+\mathbf{\Lambda }:\nabla \mathbf{u}+a{\rho ^{2}}%
\nabla \cdot \mathbf{u}} \\ \notag
&&-{K({\frac{1}{2}}\nabla \rho \cdot \nabla \rho \nabla \cdot \mathbf{u}%
+\rho \nabla \rho \cdot \nabla \left( {\nabla \cdot \mathbf{u}}\right)
+\nabla \rho \cdot \nabla \mathbf{u}\cdot \nabla \rho )},
\end{eqnarray}%
\begin{equation}
{C_{q}}={\frac{1}{{\rho {T^{2}}}}}\nabla \cdot \left( {q\rho T\nabla T}%
\right) .
\end{equation}%
Here $\rho $, $\mathbf{u}$, and $T$ are the local density,
velocity, and temperature, respectively. $\mathbf{\Lambda }=K\nabla \rho
\nabla \rho -K(\rho {\nabla ^{2}}\rho +{{{{\left\vert {\nabla \rho }%
\right\vert }^{2}}}/2})\mathbf{I}-[\rho T\nabla \rho \cdot \nabla (K/T)]\mathbf{I}$ is the contribution of the density gradient to pressure tensor, $\mathbf{I}$ is a unit tensor,
and $K$ is the surface tension coefficient. $P^{\mathrm{CS}}$ indicates the Carnahan--Starling equation of state
\begin{equation}
{P^{\mathrm{CS}}}=\rho T{\frac{{1+\eta +{\eta ^{2}}-{\eta ^{3}}}}{{{{(1-\eta
)}^{3}}}}}-a{\rho ^{2}},
\end{equation}%
with $\eta ={{b\rho }/4}$ , $a$ and $b$ are the attraction and repulsion parameters, respectively. It should be pointed out that the Prandtl number $\Pr ={\mu/{\kappa_T}}={\tau/{ {2(\tau  - q)}}}$ can be adjusted by modulating the parameter $q$ in the term ${C_{q}}$, here $\mu=\rho T\tau$ and ${\kappa_T}=2\rho T(\tau  - q)$ are viscosity coefficient and heat conductivity, respectively.

Under the constraint of $\int {{f^{eq}}\mathbf{\Psi }\left(
\mathbf{v}\right) }d\mathbf{v}=\sum\limits_{l}{{f_{l}}^{eq}\mathbf{\Psi }%
\left( {{\mathbf{v}_{l}}}\right) }$, Eq. (2) can be
discretized in velocity space by an appropriate Discrete Velocity Model (DVM).
Here, the D2V33 model is used, which reads
\begin{equation}
{\mathbf{v}_{0}}=0,\mathrm{\ }{\mathbf{v}_{ki}}={v_{k}}\left[ {\cos ({\frac{i%
}{4}}\pi ),\mathrm{\ }\sin ({\frac{i}{4}}\pi )}\right] ,
\end{equation}%
where $k=1,2,3,4$ is the $k$-th group of particle velocities whose
speed is ${v_{k}}$ and $i=1,\cdots ,8$ is the direction of ${v_{k}}$ , as
shown in Fig. 1.
We stress that the DVM is selected according to modeling accuracy and stability of the model.

\begin{figure}[tbp]
{\centering
\centerline{\epsfig{file=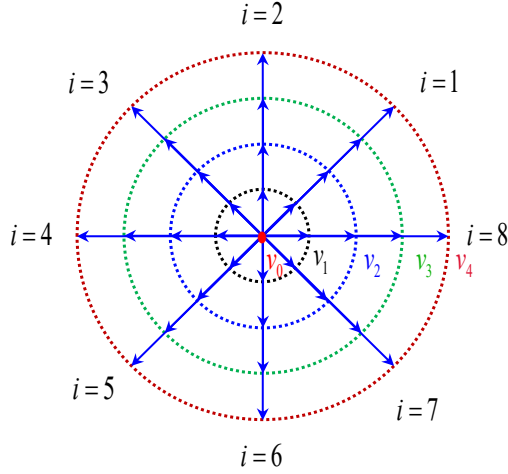,bbllx=260pt,bblly=200pt,bburx=630pt,bbury=440pt,
width=0.6\textwidth,clip=}}}
\caption{Schematic diagram of the discrete velocity model.}
\end{figure}
Then, the discrete GLS-Boltzmann equation can be written as
\begin{equation}
{\frac{{\partial {f_{ki}}}}{{\partial t}}}+{\mathbf{v}_{ki}}.{\frac{\partial
}{{\partial \mathbf{r}}}}{f_{ki}}=-{\frac{1}{\tau }}({f_{ki}}
-f_{ki}^{eq})+{I_{ki}},
\end{equation}%
where $f_{ki}^{eq}$ is the discrete version of the local equilibrium
distribution function; ${I_{ki}}$ takes the following form
\begin{equation}
{I_{ki}}=-[A+{\mathbf B}\cdot ({\mathbf v}_{ki}-{\mathbf u })+(C+{C_{q}}){({%
\mathbf v_{ki}}-{\mathbf u})^{2}}]f_{ki}^{eq},
\end{equation}%
the discrete equilibrium distribution function $f_{ki}^{eq}$ is
\begin{equation}
f_{ki}^{eq}=\rho {F_{k} [( {1-{\frac{{u^{2}}}{{2T}}}+{\frac{{u^{4}}}{{8{%
T^{2}}}}}}) +{\frac{{{\mathbf{v}_{ki}}\cdot \mathbf{u}}}{T}}( {1-{%
\frac{{u^{2}}}{{2T}}}}) +{\frac{{{{({\mathbf{v}_{ki}}\cdot \mathbf{u})}%
^{2}}}}{{2{T^{2}}}}}( {1-{\frac{{u^{2}}}{{2T}}}}) +{\frac{{{{({%
\mathbf{v}_{ki}}\cdot \mathbf{u})}^{3}}}}{{6{T^{3}}}}}+{\frac{{{{({\mathbf{v}%
_{ki}}\cdot \mathbf{u})}^{4}}}}{{24{T^{4}}}}}}],
\end{equation}%
with
\begin{equation}
{F_{1}}={\frac{{48{T^{4}}-6(v_{2}^{2}+v_{3}^{2}+v_{4}^{2}){T^{3}}%
+(v_{2}^{2}v_{3}^{2}+v_{2}^{2}v_{4}^{2}+v_{3}^{2}v_{4}^{2}){T^{2}}-{\frac{1}{%
4}}v_{2}^{2}v_{3}^{2}v_{4}^{2}T}}{{%
v_{1}^{2}(v_{1}^{2}-v_{2}^{2})(v_{1}^{2}-v_{3}^{2})(v_{1}^{2}-v_{4}^{2})}}},
\end{equation}%
\begin{equation}
{F_{2}}={\frac{{48{T^{4}}-6(v_{1}^{2}+v_{3}^{2}+v_{4}^{2}){T^{3}}%
+(v_{1}^{2}v_{3}^{2}+v_{1}^{2}v_{4}^{2}+v_{3}^{2}v_{4}^{2}){T^{2}}-{\frac{1}{%
4}}v_{1}^{2}v_{3}^{2}v_{4}^{2}T}}{{%
v_{2}^{2}(v_{2}^{2}-v_{1}^{2})(v_{2}^{2}-v_{3}^{2})(v_{2}^{2}-v_{4}^{2})}}},
\end{equation}%
\begin{equation}
{F_{3}}={\frac{{48{T^{4}}-6(v_{1}^{2}+v_{2}^{2}+v_{4}^{2}){T^{3}}%
+(v_{1}^{2}v_{2}^{2}+v_{1}^{2}v_{4}^{2}+v_{2}^{2}v_{4}^{2}){T^{2}}-{\frac{1}{%
4}}v_{1}^{2}v_{2}^{2}v_{4}^{2}T}}{{%
v_{3}^{2}(v_{3}^{2}-v_{1}^{2})(v_{3}^{2}-v_{2}^{2})(v_{1}^{2}-v_{4}^{2})}}},
\end{equation}%
\begin{equation}
{F_{4}}={\frac{{48{T^{4}}-6(v_{1}^{2}+v_{2}^{2}+v_{3}^{2}){T^{3}}%
+(v_{1}^{2}v_{2}^{2}+v_{1}^{2}v_{3}^{2}+v_{2}^{2}v_{3}^{2}){T^{2}}-{\frac{1}{%
4}}v_{1}^{2}v_{2}^{2}v_{3}^{2}T}}{{%
v_{4}^{2}(v_{4}^{2}-v_{1}^{2})(v_{4}^{2}-v_{2}^{2})(v_{4}^{2}-v_{3}^{2})}}},
\end{equation}%
\begin{equation}
{F_{0}}=1-8({F_{1}}+{F_{2}}+{F_{3}}+{F_{4}}).
\end{equation}

Taking the moments of Eq. (10) with the collision invariant vector $1$, ${\mathbf{v}_{ki}}$, and ${{{v}_{ki}^{2}}/2}$, the generalized Navier-Stokes equations for
nonideal fluid with surface tension effect are obtained \cite{RN40}
\begin{equation}
{\frac{{\partial \rho }}{{\partial t}}}+\nabla \cdot (\rho \mathbf{u})=0,
\end{equation}%
\begin{equation}
{\frac{{\partial (\rho \mathbf{u})}}{{\partial t}}}+\nabla \cdot (\rho
\mathbf{uu}+P\mathbf{I})+\nabla \cdot (\mathbf{\Lambda }+\mathbf{\Delta }%
_{2}^{\ast })=0,
\end{equation}%
\begin{equation}
{\frac{{\partial {e_{T}}}}{{\partial t}}}+\nabla \cdot ({e_{T}}\mathbf{u}+P%
\mathbf{u})+\nabla \cdot \left[ {(\mathbf{\Lambda }+\mathbf{\Delta }%
_{2}^{\ast })\cdot \mathbf{u}+\mathbf{\Delta }_{3,1}^{\ast }}\right] =0,
\end{equation}
where $\nabla \cdot \mathbf{\Lambda }$ is the surface tension \cite{RN43}, ${e_{T}}=\rho T-a{\rho ^{2}}+{{K{{\left\vert {\nabla \rho }\right\vert }^{2}}}/2+\rho u^{2}/2}$ is the total energy density.

More importantly, DBM can quantitatively provide the local TNE effects by defining thermodynamic non-equilibrium moments $\mathbf{\Delta }_{m,n}^{*}$ as
\begin{equation}
\mathbf{\Delta }_{m,n}^{\ast }=\mathbf{M}_{m,n}^{\ast }-\mathbf{M}%
_{m,n}^{\ast eq},
\end{equation}%
and
\begin{equation}
\mathbf{M}_{m,n}^{\ast }\left( {{f_{ki}}}\right) =\sum\limits_{ki}{{f_{ki}}%
\overbrace{\underbrace{({\mathbf{v}_{ki}}-\mathbf{u})({\mathbf{v}_{ki}}-%
\mathbf{u})\cdots ({\mathbf{v}_{ki}}-\mathbf{u})}_{n}{{\left\vert {({\mathbf{%
v}_{ki}}-\mathbf{u})}\right\vert }^{(m-n)}}}^{m}},
\end{equation}%
\begin{equation}
\mathbf{M}_{m,n}^{\ast eq}\left( {f_{ki}^{eq}}\right) =\sum\limits_{ki}{%
f_{ki}^{eq}\overbrace{\underbrace{({\mathbf{v}_{ki}}-\mathbf{u})({\mathbf{v}%
_{ki}}-\mathbf{u})\cdots ({\mathbf{v}_{ki}}-\mathbf{u})}_{n}{{\left\vert {({%
\mathbf{v}_{ki}}-\mathbf{u})}\right\vert }^{(m-n)}}}^{m}},
\end{equation}
$m$ is the number of velocities used in the moment and \emph{n} is the tensor
order, respectively. If $m=n$, $\mathbf{M}_{m,n}^{\ast }=%
\mathbf{M}_{m}^{\ast }$. For example%
\begin{equation}
\mathbf{\Delta }_{2}^{\ast }=\mathbf{M}_{2}^{\ast }-\mathbf{M}_{2}^{\ast
eq}=\sum\limits_{ki}{({\mathbf{v}_{ki}}-\mathbf{u})({\mathbf{v}_{ki}}-%
\mathbf{u})({f_{ki}}-f_{ki}^{eq})},
\end{equation}%
\begin{equation}
\mathbf{\Delta }_{3}^{\ast }=\mathbf{M}_{3}^{\ast }-\mathbf{M}_{3}^{\ast
eq}=\sum\limits_{ki}{({\mathbf{v}_{ki}}-\mathbf{u})({\mathbf{v}_{ki}}-%
\mathbf{u})({\mathbf{v}_{ki}}-\mathbf{u})({f_{ki}}-f_{ki}^{eq})},
\end{equation}%
\begin{equation}
\mathbf{\Delta }_{3,1}^{\ast }=\mathbf{M}_{3,1}^{\ast }-\mathbf{M}%
_{3,1}^{\ast eq}={\frac{1}{2}}\sum\limits_{ki}{{{({\mathbf{v}_{ki}}-\mathbf{u%
})}^{2}}({\mathbf{v}_{ki}}-\mathbf{u})({f_{ki}}-f_{ki}^{eq})},
\end{equation}%
\begin{equation}
\mathbf{\Delta }_{4,2}^{\ast }=\mathbf{M}_{4,2}^{\ast }-\mathbf{M}%
_{4,2}^{\ast eq}={\frac{1}{2}}\sum\limits_{ki}{{{({\mathbf{v}_{ki}}-\mathbf{u%
})}^{2}}({\mathbf{v}_{ki}}-\mathbf{u})({\mathbf{v}_{ki}}-\mathbf{u})({f_{ki}}%
-f_{ki}^{eq})}.
\end{equation}
$\mathbf{\Delta }_{2}^{\ast }$ and $\mathbf{\Delta }_{3,1}^{\ast }$ are
referred to as non-organized momentum fluxes (NOMF) and non-organized energy
fluxes (NOEF), respectively. The first-order analytical solutions for those TNE effects are given in Refs. \cite{RN26,gan2008two}
\begin{equation}
\mathbf{\Delta }_{2}^{\ast (1)}=-\rho T\tau \left[\bm{{\nabla} \mathbf{u}+{{%
(\nabla \mathbf{u})}^{T}}-\mathbf{I}\nabla \cdot \mathbf{u}}\right],
\label{eq28}
\end{equation}%
\begin{equation}
\mathbf{\Delta }_{3,1}^{\ast (1)}=-2\rho T\tau \bm{\nabla} T.  \label{eq29}
\end{equation}
It should be pointed out that the external force term is introduced into the model through $f^{eq}_{ki}$ [see Eq. (11)], so the external force term does not introduce additional first-order TNE effects compared with the ideal-gas system, but will introduce additional second-order TNE effects.

\section{Simulations and analysis}

In the simulations, the Fast Fourier Transform (FFT) scheme with $16$-th order in
precision is used to discretize the spatial derivatives; the second-order Runge-Kutta finite
difference scheme is utilized to solve the temporal derivative; the
computational grids are ${N_x} \times {N_y} = 256 \times 256$ with space
step $\mathrm{\Delta }x = \mathrm{\Delta }y = {1/128}$; the time-step is $\mathrm{\Delta }t = 0.0001$ . The parameters \emph{a} and \emph{b} in the equation of state are chosen as $a = 2.0$ and $b = 0.4$ fixing the critical point at ${T_c} = 1.88657$, ${\rho _c}
= 1.3044$, and ${P_c} = 0.8832$.

To numerically study the physical mechanisms of bubble coalescence, the
initial state of two stationary bubbles being horizontally abreast is set as
\begin{equation}
\begin{array}{l}
\rho (x,y)={\rho _{g}}+\frac{{({\rho _{l}}-{\rho _{g}})}}{2}\tanh \left[ {%
\frac{{\sqrt{{{(x-{x_{01}})}^{2}}+{{(y-{y_{01}})}^{2}}}-{r_{0}}}}{{0.5w}}}%
\right]  \\
\mathrm{\ \ \ \ \ \ \ \ \ \ \ \ \ \ \ }+\frac{{({\rho _{l}}-{\rho _{g}})}}{2}\tanh %
\left[ {\frac{{\sqrt{{{(x-{x_{02}})}^{2}}+{{(y-{y_{02}})}^{2}}}-{r_{0}}}}{{%
0.5w}}}\right] .
\end{array}
\label{eq30}
\end{equation}%
Here, ${\rho _{l}}=2.0658$, ${\rho _{g}}=0.6894$ are the density of liquid
and gas phases with $T=1.8$; $w=6.0$ is the width of boundary layer, ${r_{0}}%
=30$ is radius of the static bubble, (${x_{01}}$, ${y_{01}}$) is center
coordinates of the left bubble, (${x_{02}}$, ${y_{02}}$) is center
coordinates of the right bubble, respectively. The temperature of the system
is free to evolve during the simulations.

\subsection{Non-equilibrium characteristics of bubble coalescence}

\begin{figure}[tbp]
{\centering
\centerline{\epsfig{file=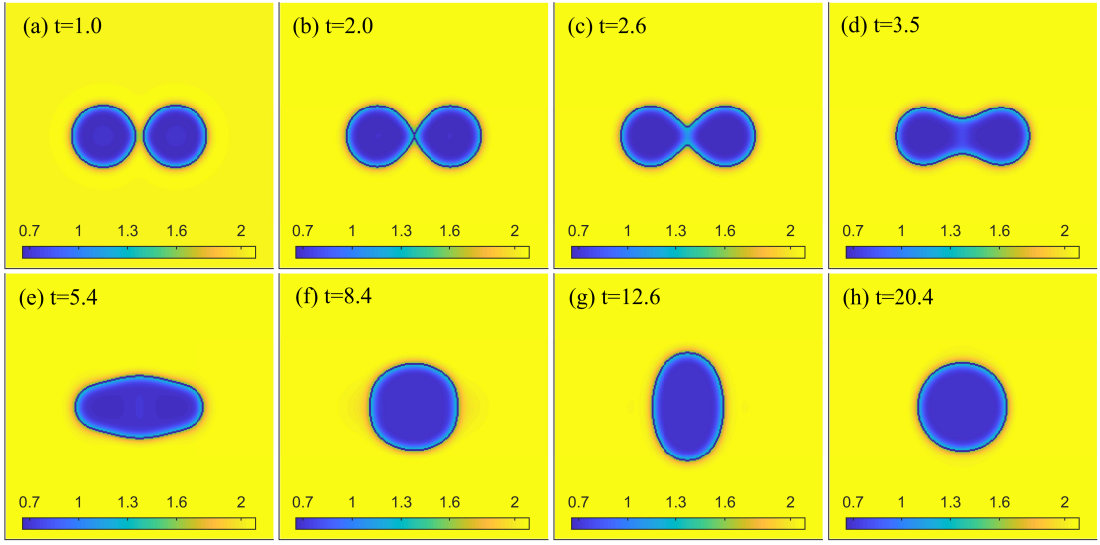,bbllx=1pt,bblly=8pt,bburx=620pt,bbury=322pt,
width=0.95\textwidth,clip=}}}
\caption{ Maps of density in the process of bubble coalescence at $
t=1.0, 2.0, 2.6, 3.5, 5.4, 8.4, 12.6$, and $20.4$ with $K=0.00015$, $\protect\tau%
=0.001$, and $\Pr=0.1$.}
\end{figure}
Two bubbles being close together will coalesce under the action of surface
tension. Figure 2 shows density patterns at eight characteristic instants with $K
= 0.00015$ and $\tau = 0.001$. Time evolutions of the ratio of
minor and major axes ${L_{d}}={{L_y}/{L_x}}$, ${L_x}$, and ${L_y}$ are displayed in Figs. 3(a) and (b), where ${L_x}$ and ${L_y}$ are the major axis and minor axis of the new bubble, respectively.
As shown in Fig. 2, the density of the
center point of the liquid film between two bubbles is less than the mean density ${\rho _c}(t) = {{({\rho _l}(t) + {\rho _g}(t))}/2}$ at $t = 2.0$; then the shape of the
coalescing bubble will go through different states:
(i) dumbbell shape at $t = 3.5$, (ii) fusiform at $t = 5.4$, (iii) unsteady circle for the first time at $t = 8.4$ with ${L_d} = 1.0$ for the first time and this instant is labeled as ${t_{l1}%
}$, as shown in Fig. 3(a), (iv) vertical ellipse at $t = 12.6$ and the ratio of minor and major axes is maximum at this time point, (v) unsteady circle for the
second time at $t = 20.4$ with ${L_d} = 1.0$ again.
From Fig. 3(b), it can
be found that the first coalescing phase ($t < {t_{l1}}$) can be divided
into two sub-stages: (a) the fast growing stage of ${L_y}$ and $L_x$ keeps almost unchanged ($t \le 5.4
$ ) and (b) the rapid decreasing stage of ${L_x}$ ($5.4 < t < 8.4$).
\begin{figure}[tbp]
{\centering
\centerline{\epsfig{file=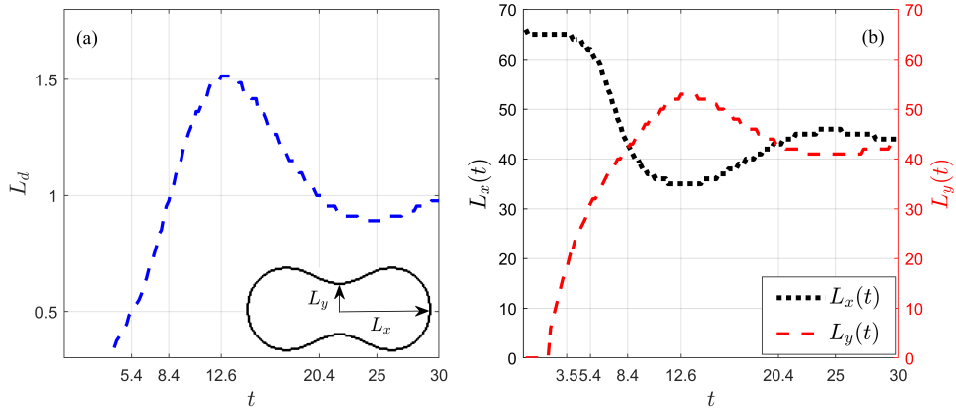,bbllx=1pt,bblly=1pt,bburx=610pt,bbury=292pt,
width=0.8\textwidth,clip=}}}
\caption{Temporal evolutions of the ratio of minor and major axes $L_d={L_y}/{L_x}$,
the major axis ${L_x}$, and the minor axis ${L_y}$ for $K=0.00015$, $\tau=0.001$, and $\Pr=0.1$.}
\end{figure}

Although the coalescent regime and the damped oscillation in the bubble
coalescence process have been extensively studied by diverse experimental,
theoretical and numerical simulation methods form different perspectives
\cite{RN10,RN12,RN14,RN15,RN16,RN17,RN18,RN23,RN24,RN25}, there are also a
lot of valuable physical problems being worth paying attention to during the
first coalescing phase ($t < {t_{l1}}$), for example all kinds of
non-equilibrium effects generate and continuously enhance in this stage.

After the instant of contact ($t = 2.0$),  an interface with negative curvature has been formed at the saddle point, together with a low temperature and pressure region near the middle liquid film, which drive mass flux from each bubble to the middle bridge. Therefore the NOMF or other non-equilibrium behavior related to mass flux must be relatively strong in the bubble. Obtained from Eq. (28), $\Delta _{2\alpha \beta }^*$ are all mainly determined by the spatial distribution of velocity gradient under the condition of viscosity coefficient being constant. As shown in Figs. 4(a) and (b), when the two bubbles are close to each other, the intermediate liquid film is thinning. The strong non-equilibrium effects first occur at the liquid film between the two bubbles because of the formation of a local relatively high velocity gradient.
\begin{figure}[tbp]
{%
\centerline{\epsfig{file=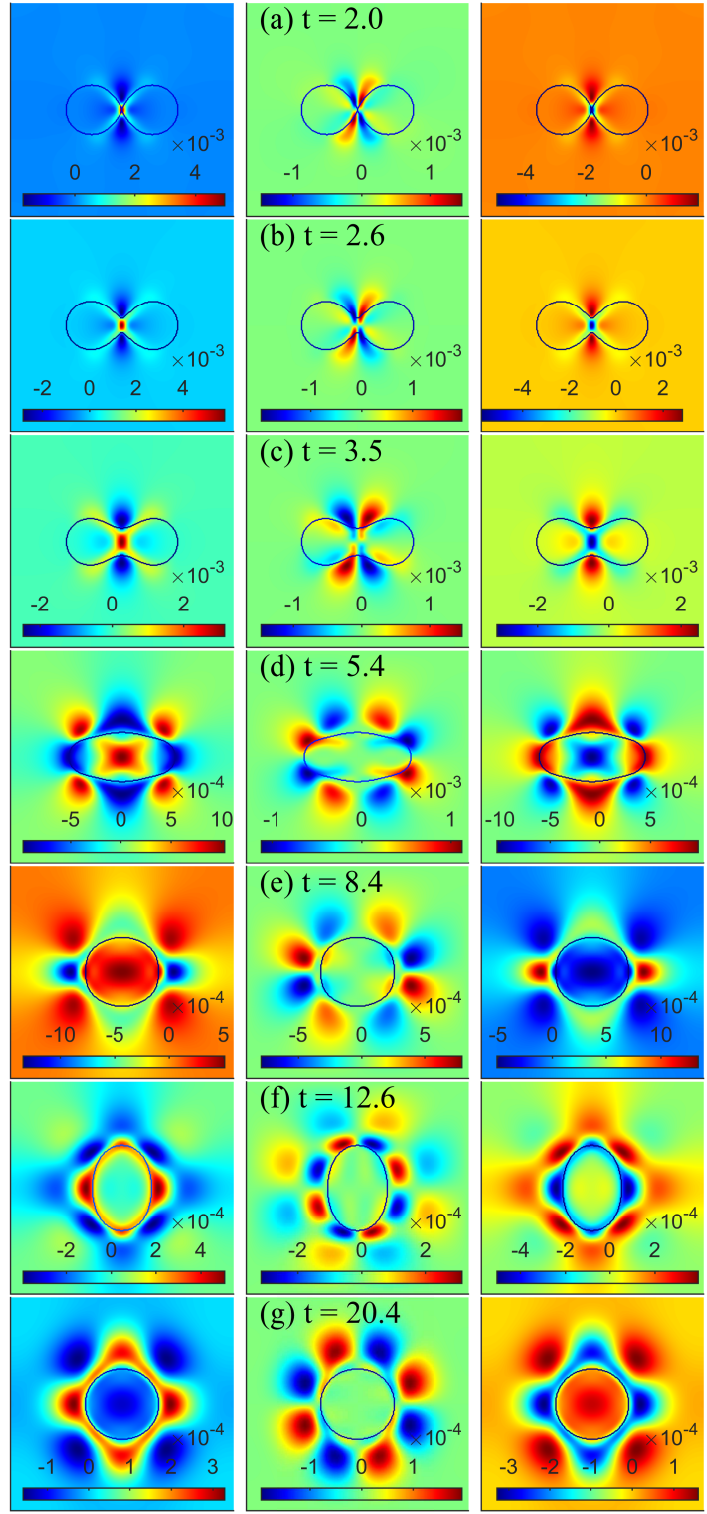,bbllx=1pt,bblly=1pt,bburx=310pt,bbury=612pt,
width=0.65\textwidth,clip=}}}
\caption{The spatial distributions of $\Delta _{2xx}^*$ (the first column), $%
\Delta _{2xy}^*$ (the second column), and $\Delta _{2yy}^*$ (the third
column) at seven characteristic moments with $K=0.00015$, $\protect\tau%
=0.001 $, and $\Pr=0.1$.}
\end{figure}
 As displayed by the black solid line and blue dash line in Fig. 5, the velocity gradient is mainly concentrated between $x=119$ and $137$, when $t \le 2.6$. What's more, four small vortices appear on both sides of the central axis under the combined influences of pressure, surface tension and viscosity [see Figs. 6(a)-(c)], thus the spatial distribution of $\Delta _{2xx}^*$ is positive in the middle and negative on both sides; the distribution of $\Delta _{2yy}^*$ is just opposite; the spatial distribution of $\Delta _{2xy}^*$ is an anti-symmetric internal and external double quadrupole structure, and the outside one is dominant. The maximum of $\Delta _{2xx}^*$ is reached soon after the merge of the two bubbles ($t = 2.6$), because surface energy rapidly translates into kinetic energy \cite{RN12}, which results in the largest velocity gradient being formed near the coalescing point, as shown in Fig. 5(b) denoted by the blue dash line. When $2.6 < t \le 5.4$, the part inside the bubble are gradually mobilized and the average velocity increases gradually, which can be illustrated by the progressively wider profiles of ${
u_{x \rm m}}$ and ${\partial {u_{x \rm m}}}/{\partial x}$ [see the pink dots line in Figs. 5(a) and (b)]. Hence the area where $\mathbf{\Delta _{2}^*}$ dominates progressively increases and its mean intensity decreases simultaneously. When $5.4 < t < 8.4$, ${L_x}$ rapidly decreases [see the black dots line in Fig. 3(b)] causing the peak of velocity to step to the outside of the bubble [see Figs. 6(e)-(f)], and then $\Delta _{2xx}^*$ in the bubble gradually becomes completely positive due to the negative velocity gradient. $\Delta _{2yy}^*$ has similar evolvement rules.

\begin{figure}[tbp]
{%
\centerline{\epsfig{file=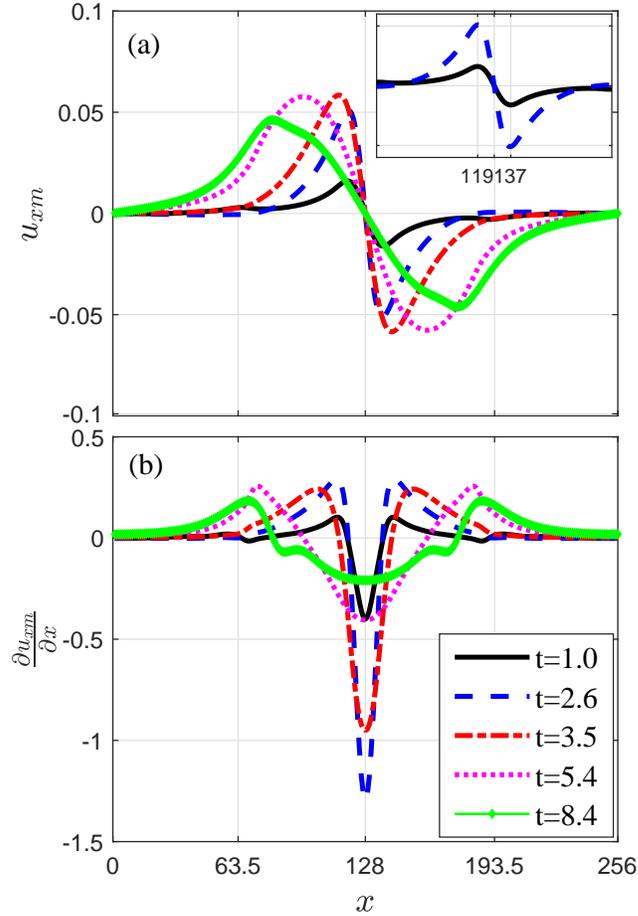,bbllx=1pt,bblly=1pt,bburx=300pt,bbury=382pt,
width=0.6\textwidth,clip=}}}
\caption{Distributions of ${u_x}$ and $ \partial u_{x \rm m}/ \partial x$
on the horizontal central axis at $t=1.0$, $2.6$, $3.5$, $5.4$, and $8.4$ with $K = 0.00015$, $\tau = 0.001$, and $\Pr = 0.1$.}
\end{figure}
For $\Delta _{2xy}^*$, when $t \le 5.4$, the relative intensity and major
area of internal quadrupole structure becomes great slowly by the gradually increasing of average speed and area of vortex flow [see the middle column in Figs. 4(a)-(c)]. The internal quadrupole structure also moves to the outside of the bubble, which leads to the formation of an outer octupole structure [see the middle column in Fig. 4(d)], because the centers of the vortex move from the saddle position to the outer of the bubble [see Figs. 6(b)-(e)]. When $t \ge 8.4$, the bubble coalescence enters the damping oscillation stage, and the spatial distribution of the three components of $\mathbf{\Delta }_2^* $ will change with the periodical variation of the vortex velocity. For example, the values of ${L_d}$ are both $1.0$ at $t = 8.4$ and $t = 20.4$ [see Fig. 3(a)], but the polarity of $\Delta _{2\alpha \beta }^*$ are all opposite [see Figs. 4 (e) and (g)], and $t = 12.6$ is the transitional moment of alternation of two polarities.
\begin{figure}[tbp]
{%
\centerline{\epsfig{file=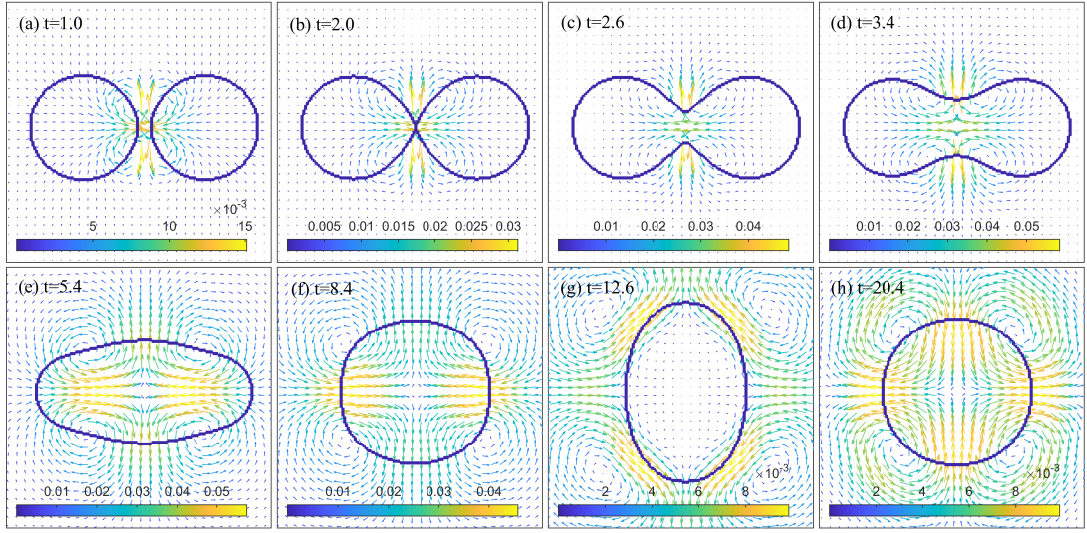,bbllx=1pt,bblly=1pt,bburx=770pt,bbury=372pt,
width=0.95\textwidth,clip=}}}
\caption{Velocity vector fields at eight characteristic moments with $%
K=0.00015$, $\protect\tau=0.001$, and $\Pr=0.1$.}
\end{figure}

To further study the relationship between TNE and the morphological or kinetics characteristics of bubble coalescence, the statistical means of four quantities are defined as follow
\begin{equation}
{\bar \Delta _{2\alpha \beta }^{*}}(t)={\frac{{\sum {\rho (x,y,t)\Delta
_{2\alpha \beta }^{\ast }(x,y,t)}}}{{\sum {\rho (x,y,t)}}}},
\end{equation}%
\begin{equation}
\bar{u}(t)={\frac{{\sum {\rho (x,y,t){u_{x}}(x,y,t)}}}{{\sum {\rho
(x,y,t)}}}},
\end{equation}
\begin{equation}
\bar D^*(t)={\frac{{\sum {\rho (x,y,t)\sqrt{\mathbf{\Delta }%
_{2}^{\ast 2}+\mathbf{\Delta }_{3}^{\ast 2}+\mathbf{\Delta }_{3,1}^{\ast 2}+ \mathbf{\Delta }_{4,2}^{\ast 2}}}}}{{\sum {\rho (x,y,t)}}}},
\end{equation}
\begin{equation}
\overline{{{(\bm \nabla \mathbf{u}:\bm \nabla \mathbf{u})}^{0.5}}}(t)={\frac{{\sum {%
\rho (x,y,t){{(\bm \nabla \mathbf{u}:\bm \nabla \mathbf{u})}^{0.5}}}}}{{\sum {\rho
(x,y,t)}}}}.\\
\end{equation}%
Here, the density-weighted statistical method \cite{RN21} is used to highlight relationship between the TNE effects and the morphological or kinetics characteristics of bubble coalescence for two reasons. One is the TNE effects of the system of multiphase flows mainly exist in the region where the macroscopic quantities have a larger gradient. Thus, the TNE effects mainly locate at the vicinity of boundary layer in the bubble coalescence system. The other is that we are concerned with the non-equilibrium effects inner the bubble; moreover, in the bubble, the density adjacent to the boundary layer is bigger than that is away from the boundary layer. The statistical areas are all in the left-half bubble.
Eq. (31) is the average of ${\Delta _{2\alpha \beta }^{*}}$. Based on the analysis of Fig. 4, $\Delta _{2yy}^{*}+\Delta _{2xx}^{*}=0$,
and $\bar \Delta _{2xy}^{*} =0$ due to its spatial antisymmetry, so it just needs to analyze the independent one $\bar \Delta_{2xx}^{*}$. Eqs. (32)-(34) are the average coalescing velocity, the mean total TNE strength, and the spatial average of the velocity gradient, respectively.
\begin{figure}[tbp]
{%
\centerline{\epsfig{file=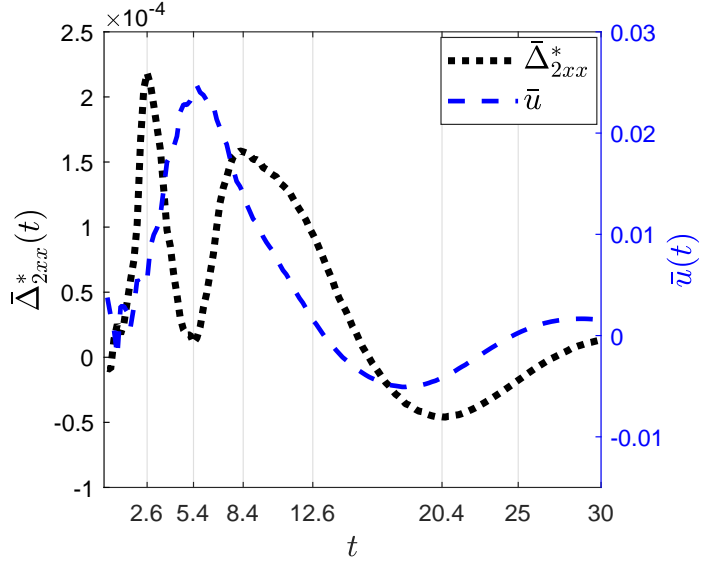,bbllx=1pt,bblly=1pt,bburx=320pt,bbury=242pt,
width=0.6\textwidth,clip=}}}
\caption{The temporal profiles of $\bar \Delta _{2xx}^*(t)$ and $%
\bar u(t)$ for $K=0.00015$, $\tau=0.001$, and $\Pr=0.1$.}
\end{figure}

Figure 7 gives time evolutions of ${\bar \Delta _{2xx}^*} (t)$ and $%
\bar u(t)$ by black dots line and blue dash line, respectively. The
evolutionary trend of the black dots line is consistent with the analysis of
Fig. 4: the value of $\bar \Delta _{2xx}^*(t)$ is increasing due to
the locally rapid growth of $\Delta _{2xx}^*$ when $t \le 2.6$; and during $2.6< t \le 5.4$, $\bar \Delta _{2xx}^*(t)$ decreases
because $\mathbf{\Delta _{2}^*}$ decreases in intensity and increases in non-equilibrium area;although the strength of non-equilibrium in the bubble is decreasing, the rate of weakening is relatively small, thus $\bar \Delta _{2xx}^*(t)$ rises up again on account of $\Delta _{2xx}^*$ inner the bubble being completely positive in the period  $5.4 < t < 8.4$; after that, it enters the damping oscillation stage ($t \ge 8.4$).

When $t \le 5.4$, ${L_y}$ rapidly increases [see the red dash line in Fig. 3(b)] while the change of ${L_x}$ is relatively small, therefore the curvature of the saddle point gradually changes from negative to positive [see Figs. 6(b)-(e)]. This process is accompanied by the release of potential energy, thus the average velocity of molecules in the bubble increases and gets the maximum at $t = 5.4$ (this instant is marked as ${t_{{\rm{umax}}}}$). When $t > t_{\rm umax}$, the average velocity diminishes gradually with the rapid decrease of ${L_x}$ and the system enters the damping oscillation stage when $t \ge 8.4$.

\begin{figure}[tbp]
{%
\centerline{\epsfig{file=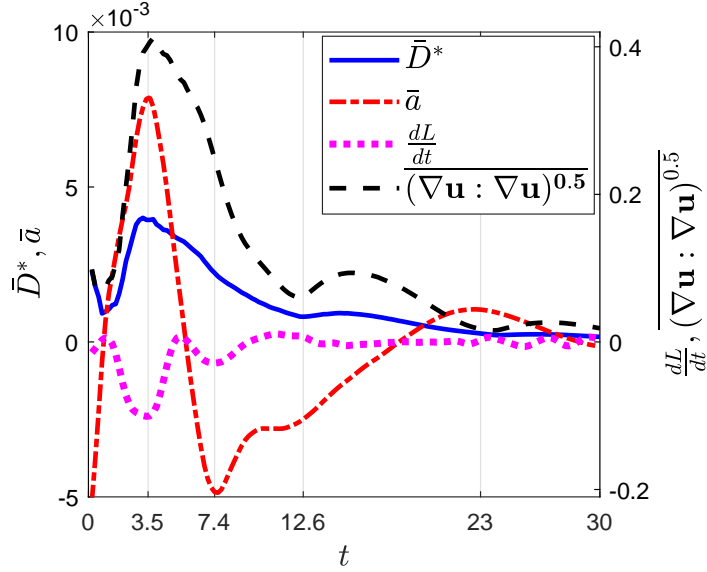,bbllx=1pt,bblly=1pt,bburx=320pt,bbury=246pt,
width=0.6\textwidth,clip=}}}
\caption{ The temporal profiles of $\bar D^*(t)$, mean acceleration of bubble coalescence $\bar{a}(t)=d\bar{u}/{dt}$, the slope of boundary length ${dL}/{dt}$, and $\overline{{{(\bm \nabla
\mathbf{u}:\bm \nabla \mathbf{u})}^{0.5}}}$ for $K=0.00015$, $\protect\tau =0.001$, and $\Pr =0.1$.}
\end{figure}
The value of $\bar D^*(t)$ mainly depends on the strength of $\mathbf{\Delta }_{2}^{*}$ and $\mathbf{\Delta }_{4,2}^{*}$, especially $\mathbf{\Delta }_{4,2}^{*}$ because its strength is about five times of $\mathbf{\Delta }_{2}^{*}$ \cite{https://doi.org/10.48550/arxiv.2203.12458}. According to the results of Ref. \cite{RN43}, $\overline{{{(\bm \nabla \mathbf{u}:\bm \nabla \mathbf{u})}^{0.5}}}$ characterizes the strength of $\mathbf{\Delta }_{2}^{*}$ and $\mathbf{\Delta }_{4,2}^{*}$. Fig. 8 demonstrates the evolutions of $\bar D^*(t)$, mean acceleration of bubble coalescence $\bar a(t) = {{d\bar u}/dt}$, the slope of boundary length $dL/dt$, and $\overline {{{(\bm \nabla \mathbf{u}:\bm \nabla \mathbf{u})}^{0.5}}} (t)$. It is interesting to find that the evolutionary trends of $\bar D^*(t)$ and $\overline {{{(\bm \nabla \mathbf{u}:\bm \nabla \mathbf{u})}^{0.5}}} (t)$ are extraordinarily similar. What is more, they are the strongest at $t = 3.5$; meanwhile, the mean coalescent acceleration is the largest and the boundary length owns the fastest changing rate. As shown in Fig. 3(b), ${L_x}$ starts to decrease at $t = 3.5$ and the morphology of the new big bubble depends on the evolutions of ${L_x}$ and ${L_y}$ thereafter, so the boundary length has the largest slop at this instant. This generates the largest energy release rate, thus $\overline {{{(\bm \nabla \mathbf{u}:\bm \nabla \mathbf{u})}^{0.5}}} (t)$ and $\bar a(t)$ reach the maximum simultaneously.

\subsection{Effects of surface tension and viscosity}

\begin{figure}[tbp]
{%
\centerline{\epsfig{file=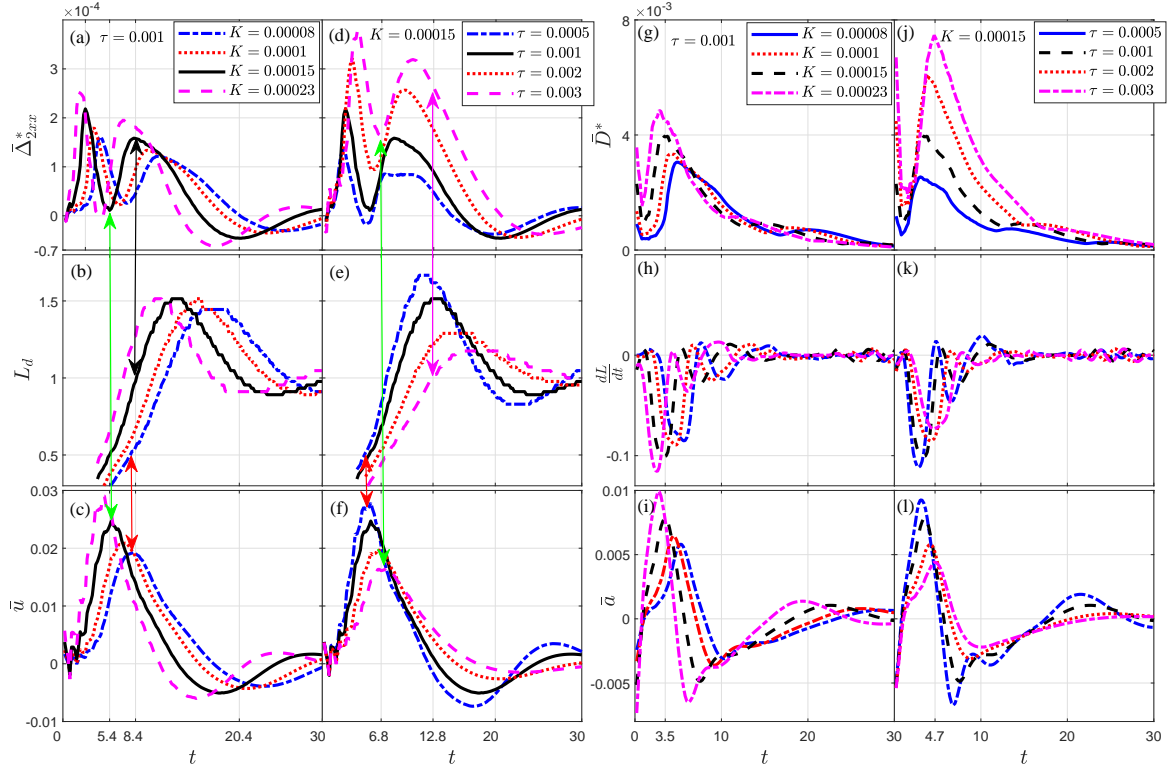,bbllx=1pt,bblly=1pt,bburx=800pt,bbury=496pt,
width=1.0\textwidth,clip=}}}
\caption{The influences of coefficient of surface tension and viscosity
coefficient. (a), (b) and (c) are profiles of $\bar \Delta _{2xx}^*(t)$, ${L_d}={L_y}/{L_x}$, and $\bar u(t)$ for $K = 0.0008$, 0.001, 0.0015, and 0.0023
with $\tau = 0.001$; (d), (e) and (f) are results for $\protect\tau %
= 0.0005$, 0.001, 0.002, and 0.003 with $K = 0.00015$; (g), (h) and (i) are
profiles of $\bar D^*(t)$, ${dL}/{dt}$, and $\bar a(t)=d \bar u/dt$ with
the same parameters with (a), (b) and (c); (j), (k) and (l) are profiles of $\bar D^*(t)$, ${dL}/{dt}$, and $\bar a(t)$ with the same
parameters with (d), (e) and (f). Here, the heat conductivity is constant
with $\tau - q = 0.005$.}
\end{figure}
\begin{figure}[tbp]
{%
\centerline{\epsfig{file=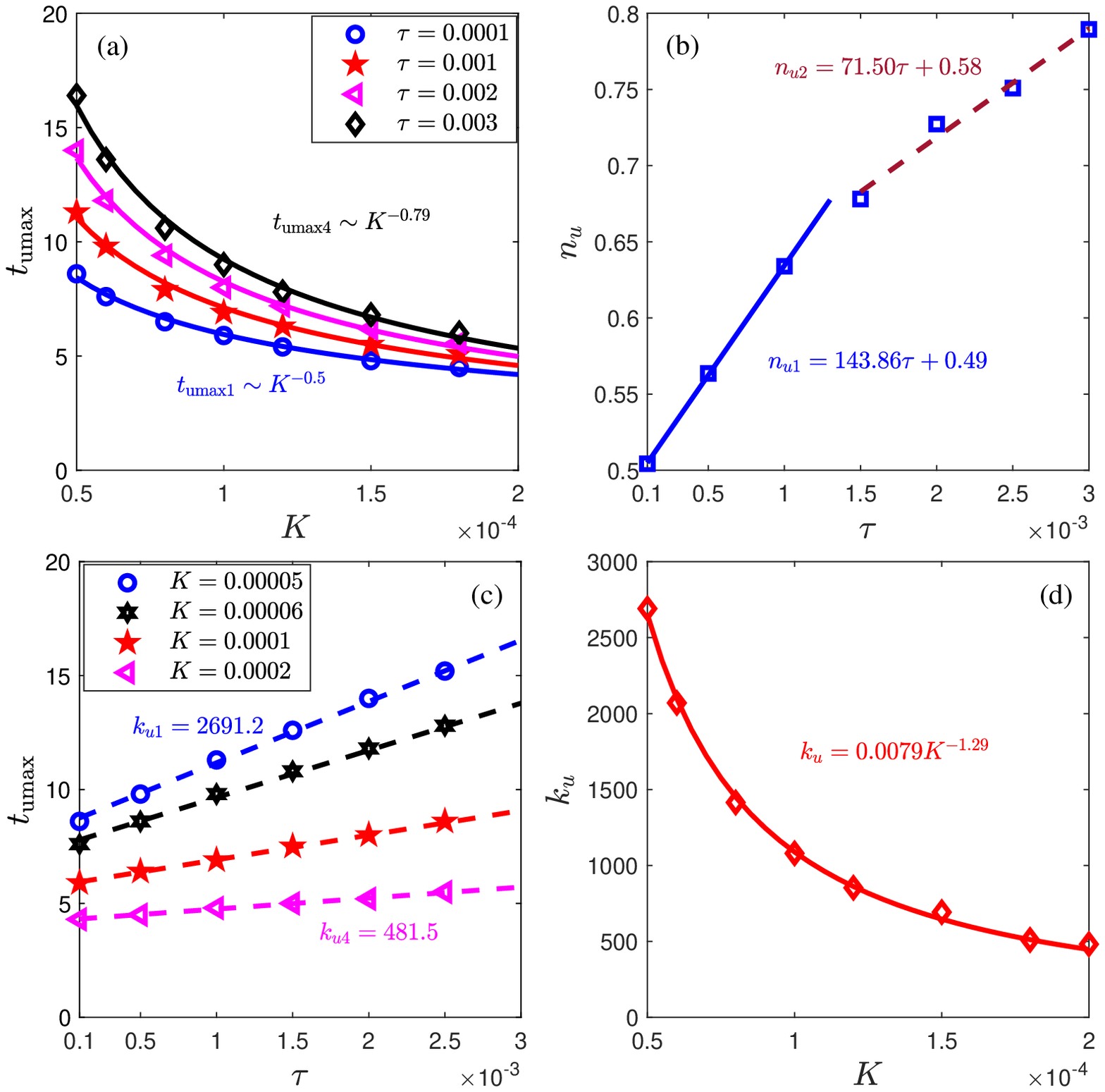,bbllx=1pt,bblly=1pt,bburx=527pt,bbury=462pt,
width=0.8\textwidth,clip=}}}
\caption{Effects of coefficient of surface tension and viscosity coefficient
on ${t_{\mathrm{umax}}}$. The relationship between ${t_{\mathrm{umax}}}$ and
coefficient of surface tension $K$ is ${t_{\mathrm{umax}}} \sim {K^{ - {n_u}}%
}$ with ${n_u} = 143.86\protect\tau + 0.49\mathrm{\ }(\protect\tau < 0.0015)$
and ${n_u} = 71.50\protect\tau + 0.58\mathrm{\ }(\protect\tau \ge 0.0015)$.
What is more, ${n_u} \sim 0.5$ for $\protect\tau = 0.0001$ and this is
consistent with the results of Ref. \cite{RN12} under the condition of lower viscosity. And there's a linear relationship between ${t_{\mathrm{umax}}}$
and viscosity coefficient ${t_{\mathrm{umax}}} = {k_u}\protect\tau + {b_u}$
with ${k_u} = 0.0079{K^{ - 1.29}}$. Here, $\protect\tau - q = 0.005$ to keep
heat conductivity being constant.}
\end{figure}
The effects of coefficient of surface tension and viscosity coefficient on the bubble coalescence are studied respectively in this section. The viscosity coefficient $\mu = \rho T\tau $ is changed by adjusting the relaxation time $\tau $. In Fig. 9, panels (a)-(f) show the evolution curves of $\bar \Delta _{2xx}^*(t)$, ${L_d}$, and $\bar u(t)$; panels (g)-(l) show the evolution curves of $\bar D^*(t)$, ${dL}/{dt}$, and $\bar a(t)$, respectively. As indicated by the green double arrows, it is clear that $\bar \Delta_{2xx}^*(t)$ of any case reaches the minimum for the first time at $t=t_{\rm umax}$. As shown by the black double arrow, $\bar \Delta_{2xx}^*(t)$ reaches its maximum for the second time at $t=t_{l1}$, i.e., the ratio of minor and major axes ${L_d}=1$ at this instant. Of particular note is that if the surface tension is relatively small or the viscosity is relatively high, $\Delta_{2xx}^*(t)$ reaches its maximum for the second time before $t=t_{l1}$, as shown by the purple double arrow. Because viscosity dominates over surface tension, which results in the relatively quick damping of non-equilibrium strength and the slower evolution speed of the system. Comprehensive statistics demonstrates that the instant that $\Delta_{2xx}^*(t)$ reaches its maximum for the second time is always equal to $t_{l1}$ when $\mathrm{Oh} \le 0.17$.
Here the Ohnesorge number $\mathrm{Oh}={\mu }/\sqrt{{\rho_{l}r_{0}\sigma }}$ \cite{RN11} is used to characterize the bubble coalescence with $\mu = \rho T\tau $, ${\rho _l}$, ${r_0}$, and $\sigma $ being viscosity, density of liquid, the initial bubble radius, and surface tension respectively. In addition, for all ${L_d}$, they are about $1/2$ at $t = {t_{\mathrm{umax}}}$, as illustrated by the red double arrows.
\begin{figure}[tbp]
{%
\centerline{\epsfig{file=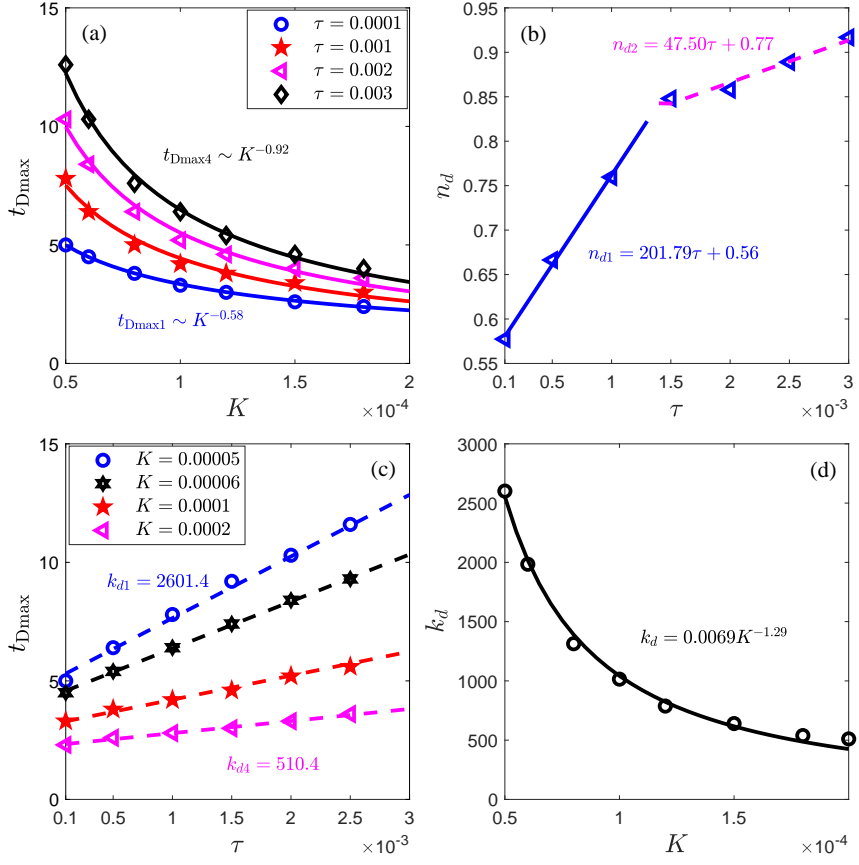,bbllx=1pt,bblly=5pt,bburx=527pt,bbury=476pt,
width=0.8\textwidth,clip=}}}
\caption{Effects of coefficient of surface tension and viscosity coefficient
on ${t_{\mathrm{Dmax}}}$ with $(\tau - q) = 0.005$.}
\end{figure}

There are three main mechanisms (surface-tension, viscous, and inertial force) influencing the process of bubble coalescence, and they dominate at different time periods \cite{RN10,RN12}. As a whole, surface tension is the original driving force of bubble coalescence and determines the
initial velocity of the saddle point, so ${t_{\mathrm{umax}}}$ decays in the form of power law as the increase of surface tension coefficient $K$, ${t_{\mathrm{umax}}} \sim {K^{ - {n_u}}}$ [see Fig. 10(a)]; viscosity impedes the coalescing progress, that is why ${t_{\mathrm{umax}}}$ linearly increases
with viscosity coefficient ${t_{\mathrm{umax}}} = {k_u}%
\tau + {b_u}$ [see Fig. 10(c)]. As shown in Fig. 10(b), the relationship between ${n_u}$ and $\tau $ is positive linear with the slope ${k_{nu1}} = 143.86$ ($\tau < 0.0015$) and ${k_{nu2}} = 71.50$ ($\tau \ge 0.0015$), i.e., viscous enhances the ${t_{\mathrm{umax}}}$ because the more viscous the liquid, the harder the displacement and the lesser the growing pre-factor of
${L_y}$ \cite{RN17,RN25}, also results in the reduction of growth rate of ${L_d}$ [see Fig. 9(e)] and the relatively large ${t_{\mathrm{umax}}}$ when surface tension coefficient is lower [see Fig. 10(a)].
And in the case of high viscosity, the hindering effect of viscosity
 on the growth of ${L_y}$ becomes stronger, as a result ${k_{nu1}} > {%
k_{nu2}}$. As exhibited in Fig. 10(d), there is a power-law fitting for the
slope ${k_u} = 0.0079{K^{ - 1.29}}$. It is quite clear that surface tension
reduces the effect of viscosity because the higher the surface tension, the
smaller the Oh number, and the bigger growing pre-factor of ${L_y}$ \cite%
{RN17,RN18,RN25}.
\begin{figure}[tbp]
{\centerline{\epsfig{file=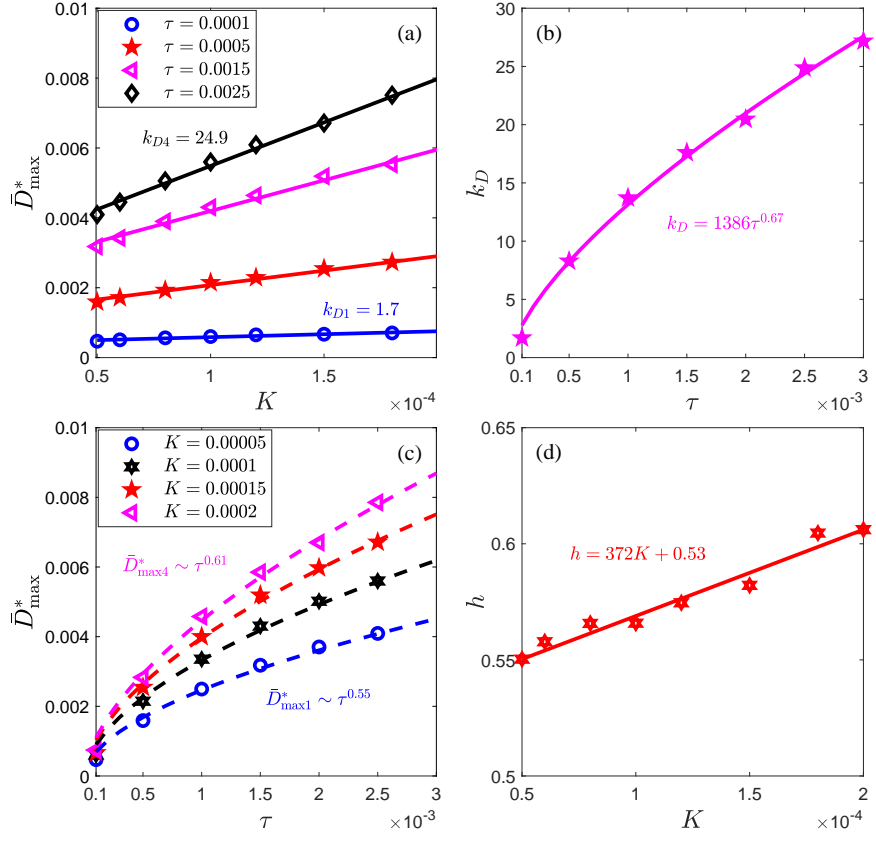,bbllx=1pt,bblly=5pt,bburx=537pt,bbury=466pt,
width=0.8\textwidth,clip=}}}
\caption{Effects of coefficient of surface tension and viscosity coefficient
on ${\bar D^* _{\mathrm{max}}}$ with $\tau - q = 0.005$.}
\end{figure}
As shown by Figs. 9 (g)-(l), the mean total TNE
strength, the average coalescent acceleration, and the absolute value of
changing rate of boundary length reach their maxima at the same time for all cases. Effects of surface tension and viscosity on ${t_{\mathrm{Dmax}}}$, which is the instant that $\bar D^*(t)$ gets the maximum, are shown in Fig. 11. On the basis of Figs. 11(a) and (b), the relation between ${t_{\mathrm{Dmax}}}$ and surface tension coefficient $K$ presents also a power law function: ${t_{\mathrm{Dmax}}} \sim {K^{ - {n_d}}}$ with ${n_d}= 201.79\tau + 0.56\mathrm{\ }(\tau < 0.0015)$ and ${n_d} = 47.50\tau + 0.77%
\mathrm{\ }(\tau \ge 0.0015)$.
As shown in Figs. 11(c) and (d), there's
a linear relation between ${t_{\mathrm{Dmax}}}$ and viscosity
coefficient: ${t_{\mathrm{Dmax}}} = {k_d}\tau + {b_d}$ with ${k_d} = 0.0069{%
K^{ - 1.29}}$. Here, the variation rules of ${t_{\mathrm{Dmax}}}$ are very
similar to those of ${t_{\mathrm{umax}}}$.

Effects of surface tension and viscosity on $\bar D^*_{\mathrm{max%
}}$, which is the maximum value of $\bar D^*(t)$, are displayed in Fig. 12.
According to Figs. 12(a) and (b), ${\bar D^*_{\mathrm{max}}}$ and $K$ shows a linear relationship ${\bar D^*_{\mathrm{max}}} = {k_D}K + {b_D}$ with ${k_D} = 1386{\tau
^{0.67}}$. From Figs. 12(c) and (d), the relation between ${%
\bar D^*_{\mathrm{max}}}$ and viscosity coefficient is a power law
function ${\bar D^*_{\mathrm{max}}} \sim {\tau ^h}$ with $h = 372K
+ 0.53$. Obviously, surface tension and viscosity are both contribute to the
growth of ${\bar D^*_{\mathrm{max}}}$, because surface tension
promotes velocity gradient $\overline {{{(\bm \nabla \mathbf{u}:\bm\nabla \mathbf{u})}^{0.5}}}$ [see Fig. 13(a)] and viscosity is the primary driving force of TNE. As shown in Fig. 13(b), $\tau $ has tripled but the peak value of $\overline {{{(\bm \nabla \mathbf{u}:\bm \nabla \mathbf{u})}^{0.5}}}$ only decreases $1.62$ times. Although the pre-factors of the first order of $\mathbf{\Delta }_{m,n}^*$ are proportional to $\tau$ \cite{https://doi.org/10.48550/arxiv.2203.12458}, our result ${\bar D^*_{\mathrm{max}}}\sim {\tau ^h} (0 < \mathrm{h} < 1)$ because the model we used essentially considers not noly the first order TNE but also the second order TNE, and the second order one is always reversed to the first order one in the two-phase flow system.
\begin{figure}[tbp]
{\centerline{\epsfig{file=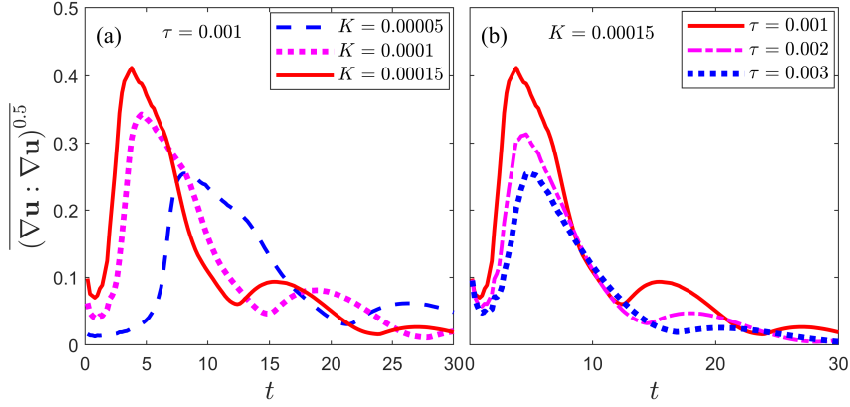,bbllx=1pt,bblly=1pt,bburx=520pt,bbury=222pt,
width=0.8\textwidth,clip=}}}
\caption{Effects of coefficient of surface tension and viscosity coefficient
on $\overline {{{(\bm \nabla \mathbf{u}:\bm \nabla \mathbf{u})}^{0.5}%
}}$ with $(\tau - q) = 0.005$.}
\end{figure}
\subsection{Effects of heat conduction}

Influences of heat conduction on ${t_{\mathrm{umax}}}$, ${t_{\mathrm{Dmax}}}$, and ${\bar D^* _{\mathrm{max}}}$ are given by Fig. 14(a).
Here, heat conductivity is changed by adjusting $\Pr $ number with fixed $\tau = 0.001$. It can be found that ${t_{\mathrm{umax}}}$, ${t_{\mathrm{Dmax}}}$, and ${\bar D^* _{\mathrm{max}}}$ are all decrease with $\tau /{\Pr}$. The relations between  ${t_{\mathrm{umax}}}$, ${t_{\mathrm{Dmax}}}$, and ${\bar D^*_{\mathrm{max}}}$ and $\tau /{\Pr}$ are ${t_{\mathrm{umax}}} = 0.038{(\tau /{\Pr})^{ - 0.66}} + 7.1$, ${t_{\mathrm{Dmax}}}= 0.028{(\tau /{\Pr})^{ - 0.75}} + 4.1$, and ${\bar D^*
_{\mathrm{max}}} = 0.0021{(\tau /{\Pr})^{ - 0.57}} + 0.00035$, respectively.
The higher the $\Pr $ number, the lower the heat conductivity. It is obvious
that the effect of heat conduction accelerates the bubble coalescence and
restrains the growth of TNE. Because the effect of heat flow is enhanced
with the increase of heat conductivity, which makes the temperature
distribution in the system more uniform. As shown in Fig. 14(b), the average
 $\overline {\left| {\bm \nabla T} \right|} $ increases with the
enhancement of $\Pr $, thus the strength of TNE positively associated
with $\bm \nabla T$ such as $\mathbf{\Delta }_3^*$ and $\mathbf{\Delta }_{3,1}^*$
 decrease with the increase of $\tau /{\Pr}$. Here $\overline{\left\vert {\bm{\nabla} T}\right\vert }={{\sum {\rho (x,y,t)\sqrt{%
{{({{\partial T}/{\partial x}})}^{2}}+{{({{\partial T/{\partial y}}})}^{2}}}}%
}/{{\sum {\rho (x,y,t)}}}}$.
\begin{figure}[tbp]
{%
\centerline{\epsfig{file=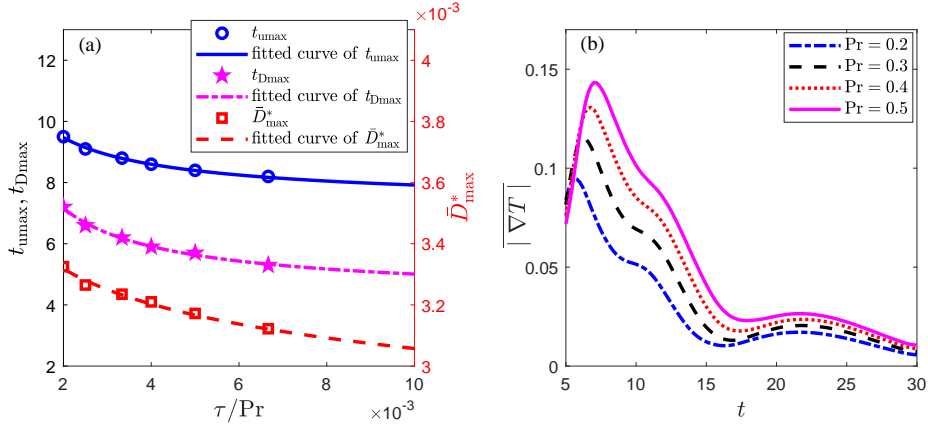,bbllx=1pt,bblly=8pt,bburx=580pt,bbury=260pt,
width=0.8\textwidth,clip=}}}
\caption{Effects of heat conductivity or $\Pr $ number. (a) effects of heat
conductivity on ${t_{\mathrm{umax}}}$, ${t_{\mathrm{Dmax}}}$, and ${%
\bar D^*_{\mathrm{max}}}$; (b) effect of $\Pr $ number on the
evolution of $\overline {\left| {\bm \nabla T} \right|}$. Here $K =
0.00008$, $\tau = 0.001$.}
\end{figure}

\section{Conclusions}
In this paper, we have studied the thermodynamic non-equilibrium effects during the coalescence of two initially motionless bubbles by a two-dimensional discrete Boltzmann model.
Our study focuses on two aspects: one is the relations among the thermodynamic
non-equilibrium inner the bubble, morphological and kinematics
characterizations of the system before the unsteady circle being formed for the first time; the other is the influences of surface tension, viscosity, and heat conduction on the average $\mathrm{\Delta }_{2xx}^*$ inner the bubble $\bar\Delta _{2xx}^*$, the instant ${t_{\mathrm{umax}}}$ that the average coalescing speed gets the maximum, the instant ${t_{\mathrm{Dmax}}}$ that the average total thermodynamic non-equilibrium strength reaches the greatest, and the greatest value of the average thermodynamic non-equilibrium strength ${\bar D^*_{\rm max}}$.

The dynamical and complex spatial distribution of $\mathrm{\Delta }_{2\alpha \beta }^*$ during bubble coalescence is investigated in detail. The strong non-equilibrium effect first occurs at the middle of two bubbles. And then the strength of non-equilibrium increases rapidly in this local area and reaches the maximum soon after the merging of the two bubbles. After that, the dominating regions of ${\Delta }_{2\alpha \beta }^*$ gradually become larger. ${\Delta }_{2xx}^*$ and ${\Delta }_{2yy}^*$ have the anti-symmetrical spatial distributions. The distribution of ${\Delta }_{2xy}^*$ changes from an anti-symmetric internal and external double quadrupole structure to an outer octupole structure. Their polarity changes with variation of the vortical direction of the velocity field periodically.

The mean value of ${\Delta _{2xx}^*}$ inner the bubble, which integrates kinematic, morphological and non-equilibrium features, can be used to calibrate the three stages of bubble coalescence. In the first stage, the minor axis grows apace until the ratio of minor and major axes is $1/2$ and the average coalescing speed gradually increases to the maximum with the rapid decrease of surface energy. In the second stage, the decreasing rate of the major axis exceeds the increasing rate of the minor axis until the ratio of minor and major axes is $1$. In the third stage, the system enters the damping oscillation stage. Due to the major axis starting to decrease, the absolute value of the slope of boundary length reaches the maximum at this time, which leads to the fastest release of surface energy, thus both the average total TNE strength in the bubble, the mean acceleration of bubble coalescence get the maximum simultaneously.

As the results of Refs \cite{RN10,RN12,RN17,RN18}, surface tension promotes the growth of the major axis while viscosity inhibits it. The relations between physical quantities that we care about (${t_{\rm umax}}$, ${t_{\rm Dmax}}$, and ${\bar D^*_{\rm max}}$) and coefficient of
surface tension and viscosity coefficient are revealed in detail. The
effect of heat conduction accelerates the bubble coalescence and restrains
the growth of the thermodynamic non-equilibrium effects, which consists with the results of Refs. \cite{RN40,RN41}
that the heat conduction facilitates the merge of small domains in the stage of domain growth. In addition, for any parameter we focused, the ratios of minor and major axes are all about $1/2$ when their average coalescence speeds reach the maxima.

The thermodynamic non-equilibrium effects present some new insights into the coalescence behavior. At the same time, the current study  presents a new perspective to detect the progress of bubble coalescence in engineering applications.

\section*{Acknowledgements}

We acknowledge support from the National Natural Science Foundation of China
(Grant Nos. 11875001, 12172061, 11974044 and 11904011), Natural
Science Foundation of Hebei Province (Grant No. A2021409001), ``Three, Three and Three''\ Talent Project of Hebei Province (Grant No. A202105005), CAEP Foundation (Grant No. CX2019033), the opening project of State Key
Laboratory of Explosion Science and Technology (Beijing Institute of Technology) (Grant No. KFJJ21-16M), Science Foundation of North China Institute of Aerospace Engineering (Grant No. KY202003).

\bibliography{Refs4}

\end{document}